\documentclass[twocolumn,showpacs,preprintnumbers,amsmath,amssymb]{revtex4}
\usepackage{epsfig}
\preprint{LCMOB/2005}

\begin{document}

\title{Infrared observation of the Hund's mechanism in an electron-doped manganite}
\author{A. Nucara$^1$, P. Calvani$^1$, F. Crispoldi$^1$, D. Sali$^{1}$, S. Lupi$^1$, C. Martin$^2$, and A. Maignan$^2$ } 
\affiliation {$^1$"Coherentia"-INFM and Dipartimento di Fisica, Universit\`a di Roma ``La Sapienza'', Piazzale A. Moro 2, I-00185 Roma, Italy}
\affiliation {$^2$Laboratoire CRISMAT-UMR 6508 ENSICAEN, 6, Bd. Mar\'echal Juin,
14050 Caen, France}
\date{\today}

\begin{abstract}
In the mid-infrared absorption of Sr$_{1-x}$Ce$_{x}$MnO$_{3}$ at low electron doping ($x$ = 0.05), a band at 0.3 eV is fully replaced by another one at 0.9 eV as the system becomes antiferromagnetic (AF) of type $G$. A weaker effect occurs at $x$ = 0.10 for an AF phase of type $C$. One thus directly measures the electron hopping energies for spin parallel and anti-parallel to that of the host ion. The Hund's, crystal-field, and Jahn-Teller splittings for the Mn$^{3+}$ ions in a Mn$^{4+}$ matrix, can also be derived. 
\end{abstract}

\maketitle

\section{Introduction}

The close interplay between charge dynamics and magnetism is the basic feature of the manganites A$_{1-x}$B$_{x}$MnO$_{3}$ (A = La,Nd and B = Sr,Ca). Generally speaking, these perovskites show poor conductivity in the paramagnetic (PM) phase at high $T$, metallic conduction in the ferromagnetic (FM) phase, charge localization and (often) ordering in the antiferromagnetic (AF) phase at low $T$.  The metallic FM phase is explained by the double-exchange  mechanism \cite{Zener,Anderson}, with corrections \cite{Millis} for the Jahn-Teller distortion that the oxygen octahedra around the Mn$^{4+}$ ions experience as they receive an additional electron. The polaronic nature of the carriers is confirmed by the observation of characteristic bands in the mid-infrared \cite{Nucara02,Kim,Boris,Quijada} which depend both on doping and temperature.

\begin{figure}
{\hbox{\epsfig{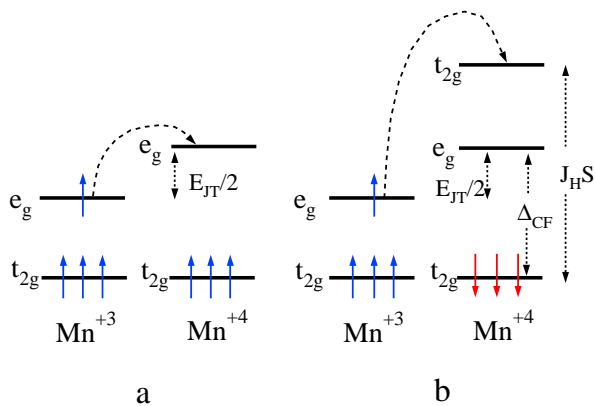}}}
\caption{(Color online) Energy diagram describing the hopping of an electron with spin parallel (a) or antiparallel (b) to that of the Mn$^{4+}$ ion. } 
\label{scheme}
\end{figure}

The insulating character of the AF phases is basically explained by the Hund's mechanism. Each Mn$^{4+}$ ion has three electrons $t_{2g}$, and spin $S$ = 3/2. An Mn$^{3+}$ ion at site $i$ has an additional electron $e_{g}$ with spin $s$ = 1/2 and Hund's energy $-J_H \vec s \cdot \vec S_i = -(1/2) J_H S$. Let $\vec S_i$ be up and the system slightly electron-doped, so that the 6 nearest neighbors at sites $j$ are all Mn$^{4+}$ with $S_j = S$. In the high-temperature PM phase the $\vec S_j$ are randomly oriented. One at least will be up, and will allow for a $i-j$ hopping with no change in the Hund's energy (Fig.\ \ref{scheme}-a). As the $j$ site is initially undistorted, the electron will then pay an energy

\begin{equation}
E_{\uparrow \uparrow} = E_{JT}/2 \, ,
\label{jt}
\end{equation}

\noindent
where $E_{JT}$ is the $e_g - e_g$ Jahn-Teller splitting. Its value may include corrections for the "breathing" distortion of the oxygen octahedra around Mn$^{4+}$ \cite{Ahn,Fratini}, or for the binding energy of a magnetic polaron which forms around the hopping electron \cite{Meskine}.

In an AF phase of type $G$ all the $\vec S_j$ are antiparallel to $\vec s$. If the system enters such phase below the N\'eel temperature $T_N$, the Hund's final energy becomes $+(1/2) J_H S$ and the hopping electron will pay an extra-energy $J_H S$.  According to calculations for CaMnO$_3$, based on the Local Density Approximation\cite{Satpathy}, the lowest antiparallel state is  $t_{2g}$. In this case the energy diagram is that of Fig.\ \ref{scheme}-b and 

\begin{equation}
E_{\uparrow \downarrow} = E_{JT}/2 + J_H S - \Delta_{CF} \, .
\label{t2g}
\end{equation}

\noindent 
Here $J_H S$ is the Hund's splitting between two $t_{2g}$ states with antiparallel spin and $\Delta_{CF}$ is the $t_{2g}$ - $e_g$ crystal-field splitting at $j$. If instead the lowest Mn$^{4+}$ state with antiparallel spin were $e_g$, the hopping energy would simply be $E_{\uparrow \downarrow} = E_{JT}/2 + J_H S$. 

As $E_{\uparrow \downarrow}$ is much larger than the thermal energy available below $T_N$, the $e_g$ electron will remain localized at $i$.   
In an AF phase of type $C$ on the other hand, both $\vec S_j$ above and below $\vec S_i$ are up while the four in-plane ones are down \cite{Dagotto}. The electron will pay just $E_{\uparrow \uparrow}$ to move along the $c$ axis, a much higher $E_{\uparrow \downarrow}$ to move in the plane. Nevertheless, it has been shown that an AF phase of type $C$ does not provides a one-dimensional metal in manganites, due to correlation effects \cite{Fang}. The opposite situation is encountered in an AF phase of type $A$, where the out-of-plane spins are parallel to $S_i$ and the in-plane ones are antiparallel. The FM planes of the $A$ phases, in manganites are close to a metallic planar instability\cite{Dagotto}.

According to the band calculations of Ref. \onlinecite{Satpathy}, in the 
all-Mn$^{4+}$ system CaMnO$_{3}$ $J_H S$ varies between 1.7 and 2.8 eV throughout the Brillouin zone. Up to now, experimental evaluations of $J_H S$ in the FM or PM phases of hole-doped manganites were extracted from bands in the visible or UV range and vary from 0.9 through 3.4 eV \cite{Okimoto,Jung1,Quijada}. In the orbital-ordered, $AF$ phase of the all-Mn$^{3+}$ system LaMnO$_{3}$, $(1/5) J_H S$ is reported to be 0.5 eV \cite{Kovaleva}. 

Here we report a direct determination of $E_{\uparrow \uparrow}$ and $E_{\uparrow \downarrow}$ based on the abrupt transition, at $T_N$, from a regime of spin-parallel hopping to one of anti-parallel hopping. We show that, in a manganite at low electron dilution, this induces a spectacular effect in its mid-infrared absorption. This effect allows one to determine, with unprecedented precision, most of the energies involved in the above Equations. The experiment is performed on Sr$_{1-x}$Ce$_{x}$MnO$_{3}$ (SCMO), where  Ce is in the +3 state \cite{Raveau} and provides the electrons, because: i) unlike the La-based manganites \cite{Hemberger}, it exhibits a good chemical stability at low doping; ii) it exhibits different AF phases and N\'eel temperatures for different $x$, and this will provide a good check of the results. 

\section{Experiment}

Two polycrystalline,  SCMO pellets have been studied here, having $x$ = 0.05 and $x$ = 0.10. For $x$ = 0.05, room-temperature x-ray diffraction data are well fitted by the cubic Pm3m space group with lattice constant $a$ = 0.38107 nm. For $x$ = 0.10 the structure is tetragonal (I4/mcm) with $a = b$ = 0.53637 nm and $c$ = 0.77481 nm. Neither sample exhibits any trace of spurious hexagonal phases and only a small volume fraction  (4 \%) of the $x$ = 0.10 sample is cubic, from pure SrMnO$_{3}$. This will not affect the optical measurements presented here, which will be focused on electronic transitions from $e_g$ states, not occupied at the $x$ = 0 composition.  

\begin{figure}
{\hbox{\epsfig{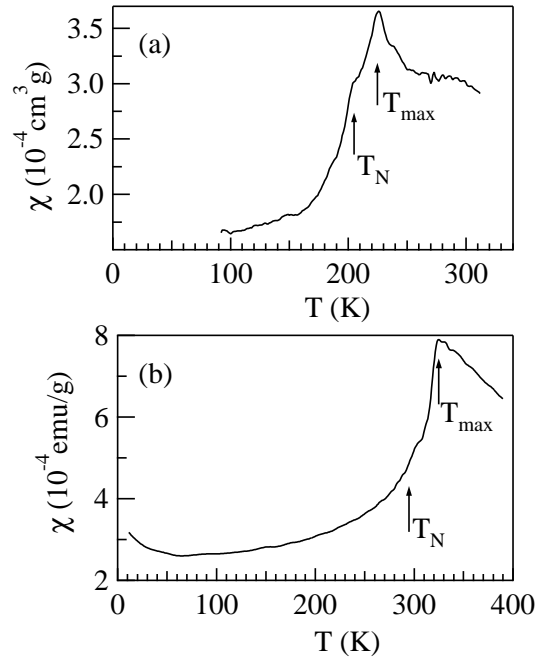}}}
\caption{Magnetic susceptibility vs. $T$ of Sr$_{0.95}$Ce$_{0.05}$MnO$_{3}$ (a) and of Sr$_{0.90}$Ce$_{0.10}$MnO$_{3}$ (b). Neutron scattering data (see text) show that the sample in a) enters an AF phase of type $G$ below $T_N \simeq$ 205 K, the one in b) an AF phase of type $C$ below $T_N \simeq$ 295 K.}
\label{chi}
\end{figure}

The magnetic susceptibility $\chi (T)$ is shown for both samples in Fig.\ \ref{chi}. Sharp maxima are found at $T_{max}$ = 225 K and 325 K for $x$ = 0.05 and 0.10, respectively. The peak at 325 K is reported to be observed also in Ref. \onlinecite{Mandal} for the same $x$ value, even if data are not shown.  A shoulder appears in both samples at $T_{sh} < T_{max}$. Neutron scattering data show that the AF order is fully established below $T_{sh}$, which therefore is the effective N\'eel temperature \cite{Raveau}. In our samples, $T_N$ is then 205 K for $x$ = 0.05,  295 K for $x$ = 0.10. The AF phase is of type $G$ \cite{Martin} at $x$ = 0.05 (as at $x$ = 0), of type $C$ \cite{Jirak} at $x$ = 0.10. This behavior is fully consistent with the general properties of manganites at high divalent dopant concentration (see, \textit{e. g.}, Fig. 1 of Ref. \onlinecite{Tobe}). Between $T_{max}$ and $T_N$ there should be a mixed AF-FM phase, as found in Ca$_{0.82}$Bi$_{0.18}$MnO$_{3}$  in correspondence of a behavior for the magnetic susceptibility \cite{Bao} quite similar to that of Fig.\ \ref{chi}.

We have measured, for both $x$ = 0.05 and 0.10, the reflectivity $R(\omega)$ of polycrystalline pellets of Sr$_{1-x}$Ce$_{x}$MnO$_{3}$,  prepared and controlled as described in Ref. \onlinecite{Raveau}. $R(\omega)$ was measured at nearly normal incidence (8$^0$) on an accurately polished surface. The absence of "ghost" peaks in $R(\omega )$ around 0.1 eV, as those reported for certain single crystals after polishing \cite{Takenaka}, excludes appreciable surface damage to our polycrystalline pellets, where the radiation penetrates deeply due to their poor optical conductivity ($\approx$ 10$^2$ $\Omega^{-1}$ cm$^{-1}$ at all frequencies and temperatures, see below). In order to minimize errors due to residual irregularities of the surface, the reference was obtained  by evaporating a metal film onto the sample (gold for $\omega <$ 14000 cm$^{-1}$, silver between 14000 cm$^{-1}$ and 20000 cm$^{-1}$). The results were then corrected for their real reflectivity. Data were collected by a rapid-scanning interferometer between 30 and 20000 cm$^{-1}$ and by
thermoregulating the samples within $\pm $ 2 K between 380 and 15 K. From 20000 to 40000 cm$^{-1}$, $R(\omega )$ was measured at room temperature with respect to a silver mirror by using a monochromator coupled to a charge-coupled device. The polycrystalline nature of the material may be source of errors, especially in the far infrared, if the transport properties are strongly
anisotropic \cite{Ortolani05}. However, in the present case, the in-plane Mn-O(2) bond length differs from the orthogonal Mn-O(1) one by less than 1.5 \% \cite{Raveau}. Moreover, anisotropies in $\sigma (\omega)$, as those reported for SCMO in Ref. \onlinecite{Tobe} above $x$ = 0.30, are rigorously excluded for our sample with $x$ = 0.05, where also the $AF$ phase is isotropic ($G$). As the same bands are observed at $x$ = 0.10, one can conclude that the latter spectra are also fully reliable. 

\section{Results and discussion}

\begin{figure}
{\hbox{\epsfig{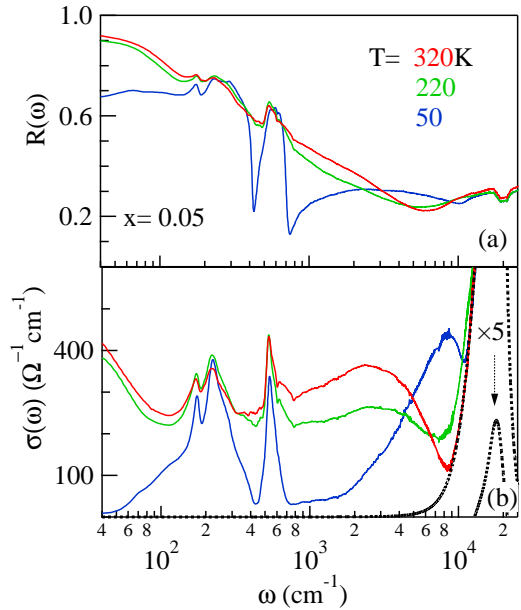}}}
\caption{(Color online) Reflectivity (a) and optical conductivity (b) of Sr$_{0.95}$Ce$_{0.05}$MnO$_{3}$. }
\label{sigma_005}
\end{figure}

Both the reflectivity $R(\omega )$ and the resulting optical conductivity $\sigma (\omega)$ are shown in Fig.\ \ref{sigma_005} for the 0.05 sample. Three main phonon lines, as predicted for the Pm3m structure of these perovskites \cite{Raveau}, are observed at low $\omega$. They are partially shielded at high $T$ by a weak Drude contribution accounting for the poor dc conductivity of the PM phase. On the opposite side, the strong band at 18000 cm$^{-1}$ is due to O-Mn charge transfer \cite{Loshka}.

\begin{figure}
{\hbox{\epsfig{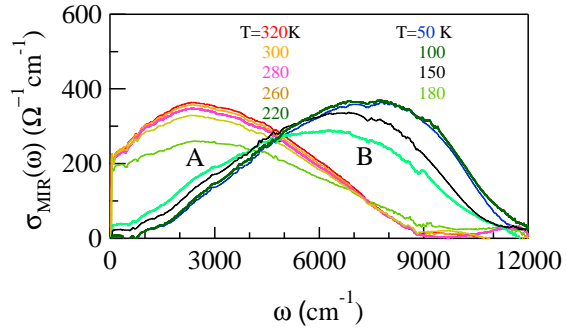}}}
\caption{ (Color online) The optical conductivity of Sr$_{0.95}$Ce$_{0.05}$MnO$_{3}$ is subtracted of the contributions in the far infrared (Drude, phonon) and in the visible range, and reported at all temperatures. Below $T_N = 205$ K the $A$ band at 0.3 eV is fully replaced by the $B$ band at 0.9 eV.}
\label{bands_005}
\end{figure} 

\begin{figure}
{\hbox{\epsfig{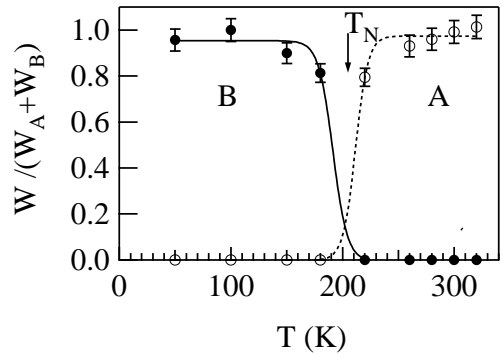}}}
\caption{Relative spectral weight vs. $T$, in Sr$_{0.95}$Ce$_{0.05}$MnO$_{3}$, of the $A$ (circles) and $B$ band (dots) of Fig. 4. The lines are guides to the eye and the arrow indicates the N\'eel temperature extracted from Fig. 2.}
\label{int_005}
\end{figure}

The usual sum rule on $\sigma (\omega)$ is fulfilled, provided that it is integrated up to 15000 cm$^{-1}$. However, in the mid-infrared range a broad band dramatically shifts towards higher energies when cooling the sample. In order to better study this effect, the imaginary part of the dielectric function $\epsilon_2 (\omega) = (4 \pi/\omega) \sigma (\omega)$ was fitted to a Drude-Lorentz model as in Ref. \onlinecite{Nucara02}. Then, both the far-infrared contributions from phonons and Drude, and the tail of the strong band at high energy in Fig.\ \ref{sigma_005} -b, were subtracted. In the resulting mid-infrared conductivity, reported in Fig.\ \ref{bands_005}, Sr$_{0.95}$Ce$_{0.05}$MnO$_3$ shows in the PM phase a single band $A$ peaked at 2500 cm$^{-1}$, or about 0.3 eV. At 220 K, as $T_N$ is approached by cooling the sample, the peak energy does not change but its intensity decreases rapidly until, at 180 K, the $A$ band is fully replaced by a $B$ band peaked at $\sim$ 7500 cm$^{-1}$, or approximately 0.9 eV. The spectral weights $W_A$ and $W_B$, obtained  by integrating the $\sigma (\omega)$ of the corresponding bands, are plotted vs. temperature in Fig.\ \ref{int_005}. Therein, one can appreciate the on-off transfer of spectral weight from $A$ to $B$, and check that it is triggered by the PM-AF transition at $T_N$.

Basing on the scenario described above, one can reasonably associate the $A$ band of the PM phase with inter-site transitions with $\vec S_i$ and $\vec S_j$ parallel. $E_{\uparrow \uparrow}$ in Eq. \ \ref{jt} is therefore 0.3 eV. The $B$ band observed below $T_N$ is instead due to the same transitions with $\vec S_i$ and $\vec S_j$ antiparallel, so that $E_{\uparrow \downarrow}$ in Eq. \ \ref{t2g} is 0.9 eV. The fact that antiparallel hopping is not observed at all in the PM phase implies that therein the $e_g$ electron tends to align the core spins of the neighboring Mn$^{4+}$ ions, forming magnetic polarons. This is indeed predicted to occur \cite{Meskine} in the electron-doping region of La$_{1-x}$Ca$_{x}$MnO$_3$ with $x \alt 1$, which corresponds to the present case. Their binding energy \cite{Meskine} of $\sim$ 0.1 eV should be subtracted to the peak energy of $A$, thus providing in Eq. \ \ref{jt} $E_{JT}/2 \sim$ 0.2 eV for a Mn$^{3+}$ ion in a matrix of Mn$^{4+}$. 

\begin{figure}
{\hbox{\epsfig{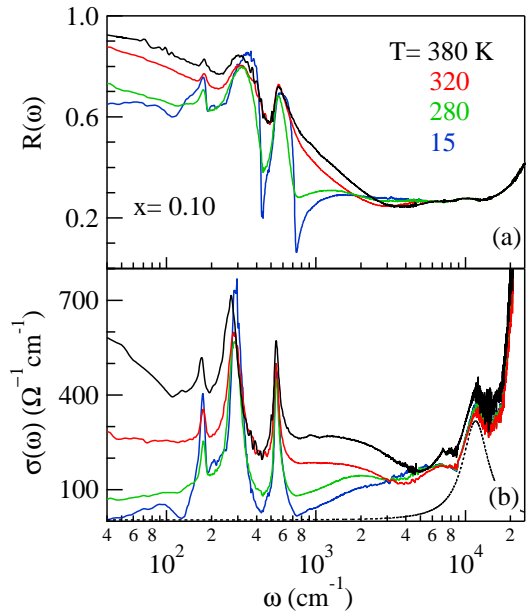}}}
\caption{(Color online) Reflectivity (a) and optical conductivity (b) of Sr$_{0.90}$Ce$_{0.10}$MnO$_{3}$  at four temperatures. In b), the 1.5 eV band is shown as provided by a Lorentzian fit.}
\label{sigma_010}
\end{figure}

The optical response of the $x$ = 0.10 sample is shown in Fig.\ \ref{sigma_010}-a and -b in terms of $R(\omega )$ and $\sigma (\omega)$, respectively, up to 380 K, well in the PM phase. Besides the phonon lines, a pronounced peak appears in $\sigma (\omega)$ at 12000 cm$^{-1}$ or 1.5 eV. Its shape is also shown as provided by a Lorentzian fit. This contribution, which at $x$ = 0.05 is not resolved from the O-Mn charge-transfer but is required by the fit, is also clearly seen in SrMnO$_3$ \cite{Sacchetti}. Bands at 1.5 eV were observed in several hole-doped manganites, and assigned to $e_g - e_g$ transitions of different kind according to different authors. They were proposed to occur within the Mn$^{3+}$ ions \cite{Jung2}, or in inter-site jumps with either $\vec s$ and $\vec S_j$ antiparallel \cite{Okimoto} or parallel \cite{Quijada}). However, the band at 1.5 eV here appears at low or even zero doping \cite{Sacchetti}, where there are no Mn$^{3+}$ ions. Moreover, 1.5 eV appears to be much higher than the expected Jahn-Teller splitting $e_g - e_g$, which should be on the scale of the lattice excitations. Therefore, also in agreement with Ref. \onlinecite{Kovaleva}, it seems reasonable to assign the 1.5 eV band to the $t_{2g} - e_{g}$ crystal field splitting $\Delta_{CF}$ in Eq.\  \ref{t2g}.

\begin{figure}
{\hbox{\epsfig{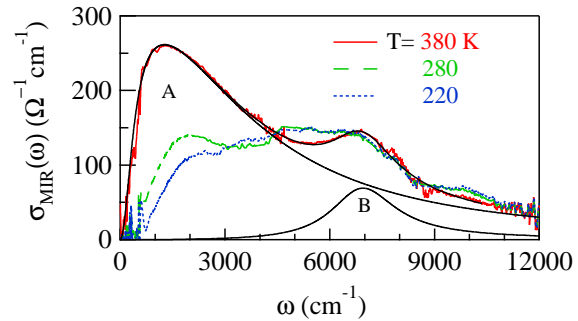}}}
\caption{(Color online) The optical conductivity of Sr$_{0.90}$Ce$_{0.10}$MnO$_{3}$  is subtracted of the contributions in the far infrared (Drude, phonon) and in the visible range. In the resulting mid-infrared absorption, a moderate transfer of spectral weight $W$ from $A$ to $B$ is evident below $T_N = 295$ K. Lorentzian fits to both bands $A$ and $B$ (see text) are shown for $T$ = 380 K.}
\label{bands_010}
\end{figure}

The subtraction procedure described above, when applied to the mid-infrared absorption of $x$ = 0.10 (Fig.\ \ref{bands_010}) shows again two bands $A$ and $B$, with $A$ which transfers most of its spectral weight to $B$ below $T_N$. Therefore, as in the 0.05 sample, magnetic polarons should be present in the PM phase at $x$ 0.10. They are expected to have a lower binding energy than at 0.05. Indeed, magnetic and dc resistivity measurements on Ca$_{1-x}$La$_{x}$MnO$_{3}$ for $x \alt 1$ show that the polaron energy may decrease by an order of magnitude as $x$ increases \cite{Neumeier}. This is fully consistent with Fig.\ \ref{bands_010}, where $A$ is peaked at 1700-2000 cm$^{-1}$ (depending on $T$). This corresponds to the bare $E_{JT}/2$ =  0.2 eV measured for $x$ = 0.05, with a negligible correction for the energy of the magnetic polaron. $B$ is peaked instead at 6500 cm$^{-1}$ or 0.8 eV. In $\epsilon_2$, both bands are well represented at any $T$ by Lorentzians which, in terms of $\sigma (\omega)$, are shown at 320 K in the same Figure. 

\begin{figure}
{\hbox{\epsfig{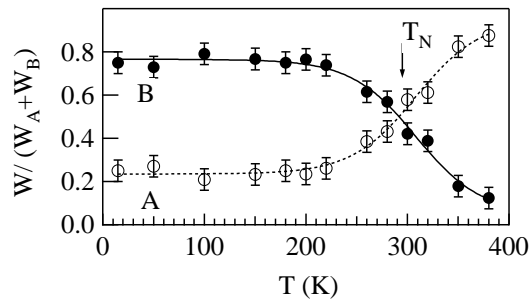}}}
\caption{Relative spectral weight vs. $T$, in Sr$_{0.90}$Ce$_{0.10}$MnO$_{3}$,  of the $A$ (circles) and $B$ bands (dots) of Fig. 7. The lines are guides to the eye and the arrow indicates the N\'eel temperature extracted from Fig. 2.}
\label{int_010}
\end{figure}

The spectral weights $W_A$ and $W_B$ change with temperature as shown in Fig.\ \ref{int_010}. A comparison with the $\chi (T)$ of Fig.\ \ref{chi}-b shows a transfer of spectral weight between the two bands around $T_N = $ 295 K, even if much less pronounced than for $x$ = 0.05. Indeed at $x$ = 0.10 both bands are present also below $T_N$, with $W_A/W_B \simeq$ 1/3. This is consistent with the 0.10 system being in a phase of type $C$ instead of $G$. Basing only on the spin population ratio in a $C$-type phase one would obtain $W_A/W_B$ = 1/2. However, the dipole matrix elements cannot be the same for $i-j$ hopping with $\vec s$ and $\vec S_j$ parallel and antiparallel, as the final states are different. Also partial charge ordering, as that observed in Nd$_{1-x}$Sr$_{x}$MnO$_{3}$ for $x >$ 0.3 \cite{Tobe}, might affect $W_A/W_B$ through an anisotropic distribution of the 10\% Mn$^{3+}$ ions. For example, if the latter ions were distributed preferentially along the $c$ axis, where the ion spins are parallel, part of the transitions which produce band $A$ would become Mn$^{3+}$ - Mn$^{3+}$. They would then move to much higher energies, due to the Hubbard repulsion, making $W_A/W_B < 1/2$.
 
One can now determine the energies involved in the electron hopping. Both $E_{JT} \sim$ 0.4 eV, on the scale of lattice excitations as expected, and $\Delta_{CF} \simeq 1.5$ eV, are measured directly. The Hund's splitting can then be found using Eq. \ref{t2g}. One finds $J_H S$ = 0.9 - 0.2 + 1.5 = 2.2  eV for $x$ = 0.05, $J_H S$ = 0.8 - 0.2 + 1.5 =  2.1 eV for $x$ = 0.10. Within the 10 \% uncertainty of the experiment both samples provide the same value, which also falls in the range calculated in Ref. \onlinecite{Satpathy} for a $t_{2g}- t_{2g}$ Hund's splitting. If instead the lowest final state with antiparallel spin were $e_g$, one would obtain $J_H S$ = 0.7 (0.6) eV  for $x$ = 0.05 (0.10). These values, however, would be much lower than usually expected for the Hund's splitting. 

In conclusion, we have directly observed the dramatic change of the mid-infrared optical conductivity induced by the onset of the Hund's mechanism at the PM-AFM transition. The effect, so clearly observed for the use of an electron system at high dilution, allowed us to determine $E_{JT}$, $\Delta_{CF}$, and $J_H S$ with good accuracy. The present study has also shown that, in a PM phase at high Mn$^{3+}$ dilution, the hopping charge can move in a locally ferromagnetic lattice, \textit{i. e.}, as a magnetic polaron. 

\acknowledgments

We wish to thank A. J. Millis and M. Grilli for useful discussions.

%
%

%
%

\end{document}